\documentclass{cta-author}
\usepackage{graphicx}
\usepackage{url}
\usepackage[caption = false]{subfig}
\usepackage{booktabs,makecell}
\usepackage{picins}
\usepackage{mwe,wrapfig}
\usepackage{tabularx}
\newcolumntype{s}{>{\hsize=0.00000005\hsize}X} 
\makeatletter
\def\hlinewd#1{%
\noalign{\ifnum0=`}\fi\hrule \@height #1 %
\futurelet\reserved@a\@xhline}
\makeatother
\usepackage{multirow}
\usepackage{amsmath}
\usepackage{amssymb}
\usepackage{enumitem}
\usepackage{array}
\usepackage{ragged2e}
\usepackage{enumitem}
\setlist{leftmargin=*,labelindent=0mm,labelsep=1.8mm}
\usepackage{float}

\begin{document}


\title{802.11 Wireless Simulation and Anomaly Detection using HMM and UBM}

\author{\au{Anisa Allahdadi$^{1}$}, \au{Ricardo Morla$^{1}$}, \au{Jaime S. Cardoso$^{1}$}}

\address{\add{1}{INESC TEC, and Faculty of Engineering, University of Porto, Porto, Portugal}
\email{anisa.allahdadi@inesctec.pt}}

\begin{abstract}
Despite the growing popularity of 802.11 wireless networks, users often suffer from connectivity problems and performance issues due to unstable radio conditions and dynamic user behavior among other reasons. Anomaly detection and distinction are in the thick of major challenges that network managers encounter. Complication of monitoring the broaden and complex WLANs, that often requires heavy instrumentation of the user devices, makes the anomaly detection analysis even harder. In this paper we exploit 802.11 access point usage data and propose an anomaly detection technique based on Hidden Markov Model (HMM) and Universal Background Model (UBM) on data that is inexpensive to obtain. We then generate a number of network anomalous scenarios in OMNeT++/INET network simulator and compare the detection outcomes with those in baseline approaches (RawData and PCA). The experimental results show the superiority of HMM and HMM-UBM models in detection precision and sensitivity.
\end{abstract}


\maketitle

\section{Introduction}

In recent years, IEEE 802.11 wireless networks has emerged as a promising technology for wireless access by mobile devices in many public places from enterprises and universities to urban areas. The flourishing popularity and ease of access to these networks has led to heavy utilization and congestion circumstances. On the other hand, interferences caused by broadcast nature of wireless links along with the other radio waves in the same frequency normally result in poor performance. In such conditions the packet transmission fails or requires several retransmission attempts causing performance issues. Furthermore, dynamic traffic loads,  evolving nature of users movement and association to different APs often induce connectivity problems in large-scale 802.11 deployments. Generally speaking, at any given moment 802.11 APs or users are likely to come across problems threatening the connection quality. 
Thus the question of performance becomes increasingly important as new applications demand sufficient bandwidth and reliable medium access. 

Across the infrastructure, there are various types of anomalous situations caused by users or APs, and automatic detection of these anomalies is of great importance for future mitigation plans. Highly utilized medium, overloaded APs, failed or crashed APs, persistent interference between adjacent APs, RF effects and authentication failure are examples of such anomalies. However, due to the time and cost limitations of constantly monitoring the entire wireless territory by sensors and sniffers \cite{Ref6,Ref11}, obtaining reliable ground truth becomes more and more challenging.

In such circumstances, when acquiring ground truth is too expensive and time-consuming, network simulations seem to be effective solutions to achieve a close to reality setup that is computationally tractable. In the research community, many wireless networks are evaluated using discrete event simulators like OMNeT++ \cite{Ref55,Ref62,Ref64}. Although having worked with other simulation frameworks such as NS3 and OPNET, we found OMNeT++/INET the most appropriate wireless network simulators for our research purposes. Besides the well-structured framework and user-friendly IDE that facilitate analysis and data gathering, OMNeT++/INET provides an adequate set of modules supporting physical and radio models for 802.11 that perfectly meet our requirements for this project.   

In our previous papers \cite{Ref23,Ref24} we utilized RADIUS authentication log data collected at the hotspot of the Faculty of Engineering of the University of Porto (FEUP). The trace data consisted of the daily summary of the connections between hundreds of APs and their corresponding wireless stations. In \cite{Ref68} we deployed a real testbed in small scale with one AP and 6 STAs using FreeRADIUS server, and generated a number of anomalies in a controlled environment for experimental purposes. In the current work we simulate a more extended WLAN with 5 APs and 30 STAs and set up several anomalous cases, including the previous ones in \cite{Ref68} and some new anomalies. We further improve our Hidden Markov Model (HMM) formerly proposed in \cite{Ref24} and \cite{Ref68} by integrating it with the concept of Universal Background Model (UBM). The simulation data are then utilized to evaluate HMM and HMM-UBM models and compare the anomaly detection results with baseline approaches (RawData and PCA). 
    
The key steps of the present work include: 1) Conducting 802.11 wireless network simulation in OMNeT++/INET to resemble normal and anomalous scenarios. 2) Reiterating the simulations with different seeds to provide miscellaneous replicates. 3) Extracting the wireless users' data, and converting it to AP usage data. 4) Building HMM and HMM-UBM models from the prepared dataset. 5) Applying the proposed anomaly detection algorithms. 6) Calculating the detection rate and sensitivity for evaluate purposes.    

Regarding the anomaly detection techniques we analyze three main approaches: 1) detection of anomalous time-series in a database of time-series, 2) recognition of anomalous patterns, and 3) detection of anomalous points within a given time-series.

Furthermore, this paper explores the following research questions: 1) whether HMM and HMM-UBM models are capable of anomaly detection and anomalous pattern recognition in AP usage data, 2) whether HMM and HMM-UBM models are required for anomaly detection or the baseline approaches are enough, 3) whether HMM-UBM have any advantages over HMM. 
 
The rest of the paper proceeds as follows. In section 2, the related work and the most recent researches relevant to the current work are presented. Section 3 characterizes the data features briefly. In section 3, the anomaly detection methodology is elaborated. Section 4 deals with the network simulation setup and focuses on the common key properties of the accomplished simulations. In section 5 the simulated scenarios are described and the experimental results are analyzed. In section 8, the main conclusions are provided and the prominent direction of future work is disclosed. 
  
\section{Related Work}

\subsection{Anomalous Pattern Detection}
In the most recent studies concerning 802.11 wireless networks, there exist several analysis on connectivity and performance issues for facilitating the network management tasks. In connection to this, a number of articles investigate the overloaded networks, faulty APs, impact of interference in chaotic 802.11 deployments and similar anomalous cases. 

Having explored the network under high medium utilization conditions, authors in \cite{Ref58} show that in the overloaded networks, stations only maintain a short association period with an AP, and repeated association and re-association attempts are common phenomena even in the absence of client mobility. Their analysis demonstrates that stations' throughput suffers drastically from the unnecessary handoffs, leading to suboptimal network performance. 

In another direction of work in \cite{Ref59}, the authors present a number of algorithms that detect failed APs by analyzing AP usage logs. The main assumption in their algorithm is that the longer the time an AP does not register events, the greater the probability that particular AP is faulty (crashed/halted). 

In relation to interference detection in WLANs, authors in \cite{Ref60} propose methods including intelligent frequency allocation across APs, load balancing of user affiliations across APs and AP adaptive power control for interference mitigation in dense 802.11 deployments. Furthermore, the authors in \cite{Ref61} studied the impact of RF interference on 802.11 networks from devices like Zigbee and cordless phones that crowd the 2.4GHz ISM band to devices like wireless camera jammers and non-compliant 802.11 devices that disrupt 802.11 operations. They affirm through practice that moving to a different channel is more effective in coping with interference than changing 802.11 operational parameters such as CCA (clear channel assessment). 

In \cite{Ref14}, a usage pattern called "abrupt ending" is explored in FEUP dataset \cite{Ref23,Ref24} that concerns the disassociation of a large number of wireless sessions in the same AP within a one second window.

The authors introduce some anomalous patterns that might be in correlation with the occurrence of this phenomena. For instance, AP halt/crash, AP overload, persistence interference and intermittent connectivity. The analysis of the anomaly-related patterns performed in this research, inspired our work to re-generate similar anomalies in network simulator in addition to the real Testbed that was already done in our previous work \cite{Ref68}. The principal goal of the simulation and the real Testbed experiments is to evaluate the HMM anomaly detection methodologies proposed in the current work as well as our former studies \cite{Ref24,Ref68}. 

\subsection{Wireless Network Simulation}

There are numerous efforts in the literature that tried to exploit simulation as an effective tool to setup a computationally tractable network. Wireless network simulation is used for various objectives from assessment and validation of models to obtain synthesized data and parameterized metrics.  
In \cite{Ref53} the authors employed simulation to generate synthetic traffic and validate their proposed model of traffic workload in a campus WLAN. As another example, the researchers in \cite{Ref54} propose a framework to integrate the infrastructure mode and ad hoc mode and they implement the framework in NS2. They used simulation to show the higher performance of their proposed model compared to the traditional wireless LAN. In a rather relevant work to ours, the performance of IEEE 802.11 wireless networks is evaluated using OPNET Modeler \cite{Ref63}. The simulated network in infrastructure mode for one AP and 12 stations investigates the performance of pure 802.11g network over a network that uses both 802.11g and 802.11b clients.

In relation to OMNet++ and its simulation models, a number of articles work on validating the reliability and accuracy of OMNeT++. For example in \cite{Ref56} the authors perform a measurement study of wireless networks in a highly controlled environment to validate the IEEE 8021.11g model of OMNet++. They used metrics like throughput, delay and packet inter-transmission to compare the measurement results to identical simulations. They show that the simulation results match the measurements well in most cases. Furthermore in \cite{Ref57} the reliability of OMNeT++ is assessed for wireless DoS attacks by comparing the simulation results to the real 802.11 testbed. In this case throughput, end-to-end delay, and packet lost ratio are considered as performance measures. The authors confirm the accuracy of the simulation results in wireless DoS domain.

However, there exist few efforts in the literature that conduct simulation of WLANs in OMNeT++ and concern about performance and quality of service (QoS). For example in \cite{Ref62} the performance of the TCP protocol for audio and video transmission is evaluated using OMNeT++ simulation. In another direction of work in \cite{Ref64} an overview of the IEEE 802.11b model is simulated in OMNeT++ and an example network consisting of a mobile station moving through a series of APs is used to analyze the handover behavior of the model. To the best of our knowledge the simulation of aforementioned anomalous patterns in WLAN infrastructure mode has never been done before. 

\subsection{HMM Applications in Network Analysis}
In wireless networking, HMMs are employed to address various aspects of network measurement and analysis. Hierarchical and Hidden Markov based techniques are analyzed in \cite{Ref15} to model 802.11b MAC-to-MAC channel behavior in terms of bit error and packet loss. The authors employed two random variables in packet loss process, inter-arrival-rate and burst-length of packet loss, and applied the traditional two-state Markov chain. The results demonstrates that two-state Markov chain provides an adequate model for the 802.11b MAC-to-MAC packet loss process. 

In \cite{Ref16} a multilevel approach involving HMMs and Mixtures of Multivariate Bernoullis (MMB) is proposed to model the long and short time scale behavior of wireless sensor network links, that is, the binary sequence of packet receptions (1s) and losses (0s) in the link. In this approach, HMM is applied to model the long-term evolution of the trace, and the short-term evolution is modeled within the states by HMM or MMB.  

In another related work, HMMs are applied for modeling and prediction of user movement in wireless networks to address QoS issues \cite{Ref18}. User movement from an AP to an adjacent AP is modeled using a second-order HMM. Although the authors demonstrated the necessity of using HMM instead of Markov chain model, the proposed model is only practical for small wireless networks with a few number of APs, not widespread WLANs.

In a more recent study in \cite{Ref72}, the authors use HMM for evaluating the performance of cooperative sensing at the fusion centre (FC). The proposed method enables the FC to become aware when the performance of cooperative spectrum sensing degrades without requiring knowledge of the local sensing statistics. Numerical results obtained from simulations confirm the effectiveness of the proposed method for both soft and hard combining schemes in practical scenarios with noise and multipath fading.

As the above literatures indicate, HMM related studies in wireless network management are rarely used specifically in performance anomaly detection of wireless networks.

\section{Data Features}

In our previous papers we utilized RADIUS authentication log data which contains session records of wireless stations connecting to APs. A preliminary analysis on the raw data yields a sequential dataset summarizing APs association history. In the current simulation we create a similar dataset with the exact same features to be synchronized with the previous HMM modeling. 
The definition of the main features along with a brief explanation on the feature selection process is presented in the following paragraphs. 

\subsection{Data Attributes}
\label{data-feat}
Data features are categorized in two main classes: \textit{Density Attributes} and \textit{Usage Attributes}. \textit{Density Attributes} demonstrate how crowded is the place in terms of active attendant users, and the \textit{Usage Attributes} disclose the volume of the sent and received traffics by the present users. The former attributes mainly characterize the association population and durability, and the later ones reveal the total bandwidth throughput regardless of how populous is the place and it is more relevant to the applications utilized by the current mobile users. 

\subsubsection{Density Attributes} 

\paragraph{User Count} the number of unique users observed in a specific location (indicated by an AP) in a time-slot. 

\paragraph{Session Count} the total population of active sessions during a time-slot regardless of the owner user. This attribute reveals the number of attempts made by the congregation of the present users to associate to the current AP. 

\paragraph{Connection Duration} the total duration of association time of all the current users. This attribute is an indicator of the overall connection persistence. The utmost amount of this feature is achieved when there is no evidence of disassociation in the ongoing active session during a time-slot.

\subsubsection{Usage Attributes} 

\paragraph{Input Data in Octets} the number of octets transmitted from the client. This attributes briefly refers to the number of bytes uploaded by the wireless user. 

\paragraph{Output Data in Octets} the number of octets received by the client. This attribute shortly refers to the number of bytes downloaded by the wireless user. 

\paragraph{Input Data in Packets} the number of packets transmitted from the client. This attribute is similar to the above \textit{Input-Octet}, just to be measured in packets.

\paragraph{Output Data in packets} the number of packets received by the client. This attribute is similar to the above \textit{Output-Octet}, just to be measured in packets.


\subsection{Feature Selection}
For subsequent analysis, we favor using less features than the entire set of attributes introduced earlier. For this purpose, we applied Principal Component Analysis (PCA) technique to find the combination of the variables which best explain the phenomena and contain the greatest part of the entire information. In the current experiment the first three principal components bring the cumulative proportion of variance to over 99\%. More detail explanation on the correlation of data features with themselves and with the principal components are provided in our previous work \cite{Ref68}. 

\section{Anomaly Detection in AP Usage Data}

We use Hidden Markov Models adapted from a Universal Background Model for 1) detection of anomalous time-series, 2) recognition of anomalous patterns, and 3) detection of anomalies within a given time-series.

\subsection{Preliminaries}

\subsubsection{Hidden Markov Model}

HMM symbolizes a doubly stochastic process with a set of observable states and a series of hidden states which can only be observed through the observable set of stochastic process. HMMs are generally used for the stochastic modeling of non-stationary time-series. 

The formal definition of a n-state HMM with Gaussian emission is determined as follows:

\noindent
\begin{itemize}
  \item A set of hidden states $S = \{s_{i}\}$	,	$1 \le i \le n$.
  \item State transition probability distribution or transition matrix. $A = \{a_{i,j}\}$ , $a_{i,j} = P(s_{j}$ at $t+1 | s_{i}$ at $t)$ , $1 \le i,j \le n$.
  \item Observation probability distributions, typically from a normal distribution in case of continuous observations. $B = \{b_{i}(o_{t})\}$ , $b_{i}(o_{t})= P(o_{t}$ at $t | s_{i}$ at $t)$ , $1 \le i \le n$.
  
 $ b_{i}(o_{t}) =  \frac{exp \{-\frac{1}{2} (o_{t}-\mu _{i})'\Sigma _{i}^{-1} (o_{t}-\mu _{i})\}}{(2\pi )^{D/2} |\Sigma _{i}|^{1/2} } $ where D refers to the dimensionality of the observation space.
  \item Initial state distribution $ \pi = \{\pi_{i}\} $,	$1 \le i \le n$ , $\pi_{i} = P(s_{i}$ at $t=1)$.
	\item $n = $ number of hidden states.
\end{itemize}

\subsubsection{Universal Background Model}

\begin{figure}
\centering
\includegraphics[width=0.49\textwidth]{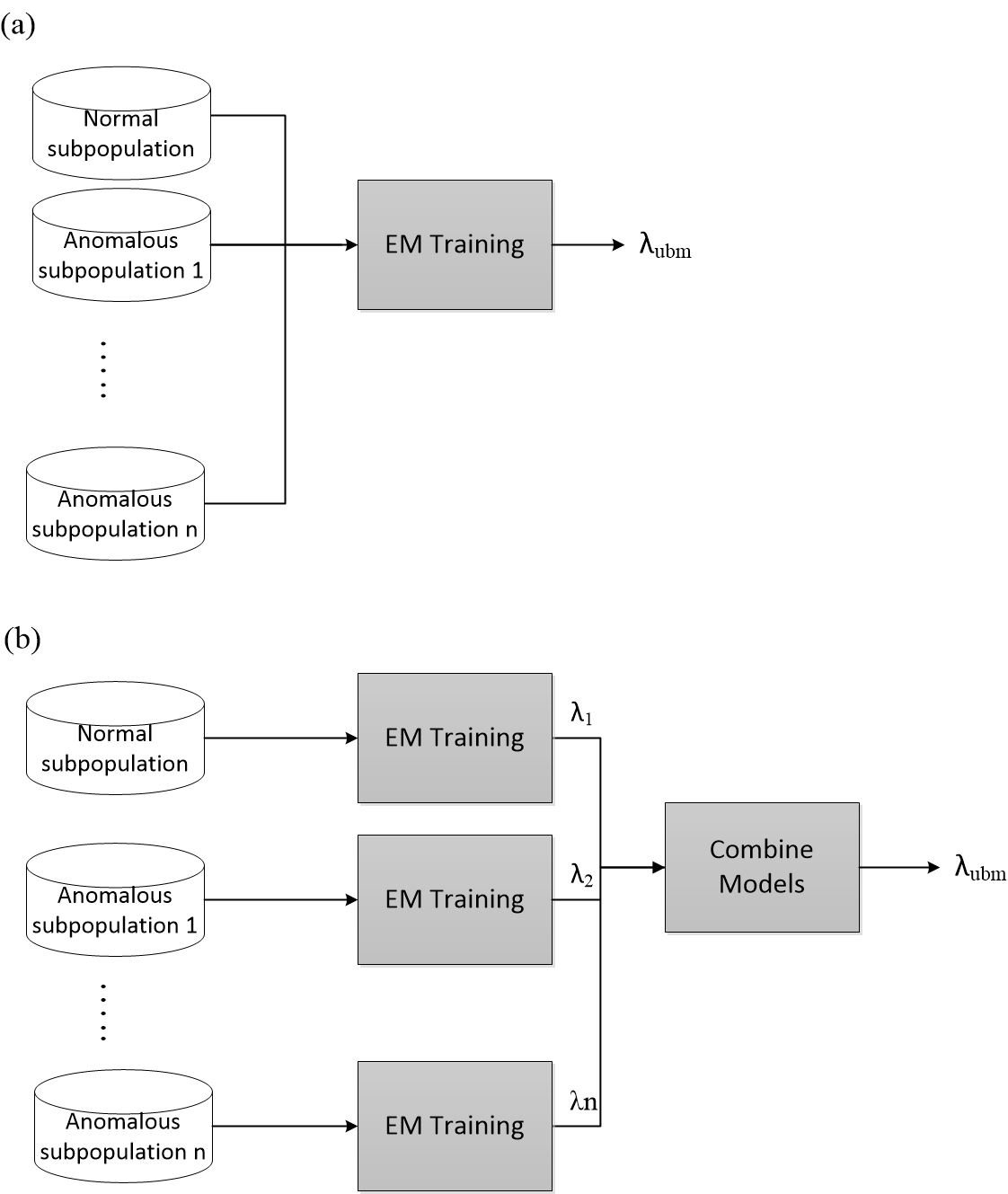}
\caption{Data and model pooling approaches for creating a UBM. (a) Data from subpopulations pooled prior to training the final UBM. (b) Individual subpopulation models trained then combined to create final UBM.}
\label{fig:20}      
\end{figure}

A universal background model (UBM) is a model used in a biometric verification system to represent general, person-independent feature characteristics to be compared against a model of person-specific feature characteristics when making an accept or reject decision. For example, in a speaker verification system, the UBM is a speaker-independent Gaussian mixture model (GMM) trained with speech samples from a large set of speakers to represent general speech characteristics. Using a speaker-specific GMM trained with speech samples from a particular enrolled speaker, a likelihood-ratio test for an unknown speech sample can be formed between the match score of the speaker-specific model and the UBM. The UBM may also be used while training the speaker-specific model by acting as the prior model in maximum a posteriori (MAP) parameter estimation \cite{Ref66}.

We applied UBM to initialize the HMM models using the data available from all experiments regardless of containing anomalies or not. This is advantageous as in unsupervised learning approach the anomalous events are not recognized antecedently. If the HMM models adapted from a UBM (HMM-UBM) produce as promising results as HMM models for labeled data, we can claim that we achieve a qualified model even in absence of the labeled data. This in turn facilitates the process of unsupervised modeling. We later compare the detection results of the HMMs initialized with and without UBM in Section \ref{anomalydetection}.  

Given the data to train a UBM, there are many approaches that can be used to obtain the final model. The simplest is to merely pool all the data to train the UBM via the EM algorithm (Figure \ref{fig:20}-a). One should be careful that the pooled data are balanced over the subpopulations within the data. Otherwise, the final model will be biased toward the dominant subpopulation \cite{Ref67}. Another approach is to train individual UBMs over the subpopulations in the data, and then pool the subpopulation models together (Figure \ref{fig:20}-b). The latter approach has the advantages that one can effectively use unbalanced data and can carefully control the composition of the final UBM \cite{Ref67}. In our model we used the first approach, and to avoid a biased model we included the same amount of normal and anomalous data sequences. Half of the dataset contains normal samples and the rest consist of anomalous events (equal portion for each anomaly).

\subsection{Anomaly Detection}

\subsubsection{Detection of anomalous time-series}

The goal of this type of anomaly detection is to find all anomalous time-series in a database of time-series, and to distinguish normal days from those which contain a number of anomalous events.  
Similar to traditional outlier\footnote{We use \textit{outlier} and \textit{anomaly} interchangeably in this context.} detection methods, the usual approach is to learn a model based on all the time-series in the database, and then compute an outlier score for each sequence with respect to the model \cite{Ref67}. In our case, we build an HMM model with UBM initialization using the training data of all the experiments. Then we calculate the log-likelihood values of each time-series in the test dataset. Those experiments that contain one or more anomalous events are expected to gain lower log-likelihood values. 

The likelihood value of HMM is the probability of an observation sequence given the model parameters. Equation \ref{equ:hmm-ll} shows how the likelihood value of HMM model $\lambda$ is calculated. 

\begin{equation}
\begin{small}
\begin{array}{r c l}
P(O|\lambda) &=& \displaystyle\sum _{all \, S} P(O|S,\lambda) P(S|\lambda) \\
             &=& \displaystyle\sum _{s_{\!1\!},s_{\!2\!},...s_{\!T\!}} \! \pi_{s_{\!1\!}}b_{s_{\!1\!}}\!(\!O_{1}\!)a_{s_{\!1\!},s_{\!2\!}}b_{s_{2}}\!(\!O_{2}\!)...a_{s_{\!T\!-\!1\!},s_{\!T\!}}b_{s_{\!T\!}}\!(\!O_{T}\!) 

\end{array}\end{small}
\label{equ:hmm-ll}
\end{equation} 

Due to the vanishingly small likelihood probabilities produced in long time-series, normally the logarithmic value is utilized.

\begin{figure}[!t]
\centering
\includegraphics[width=0.49\textwidth]{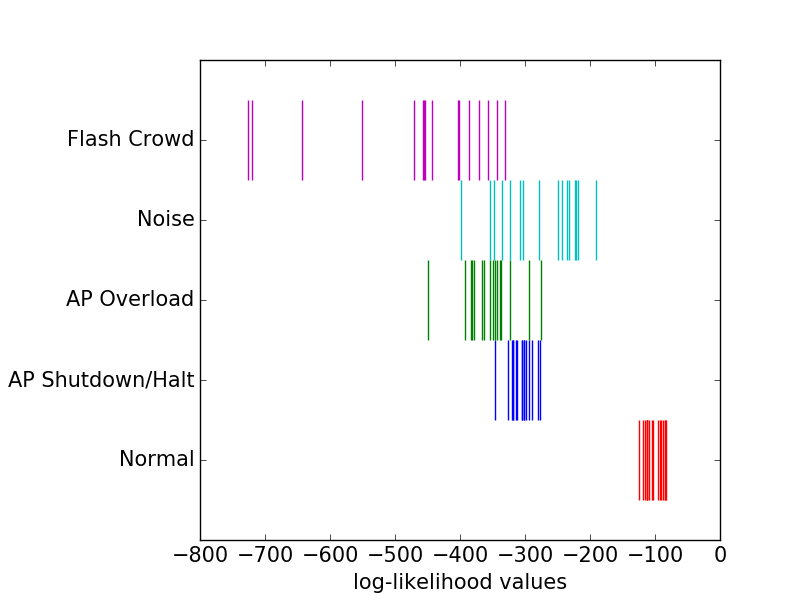}
\caption{Log-likelihood values of normal and anomalous experiments.}
\label{fig:0}      
\end{figure}

Figure \ref{fig:0} shows the range of the log-likelihood values belonging to the normal and anomalous experiments. The anomalous cases consist of \textit{AP Shutdown/Halt}, \textit{AP Overload}, \textit{Noise}, and \textit{Flash Crowd} scenarios. As this figure displays there is a distinction between the log-likelihood values of the normal cases and the rest of the anomalies. However, the anomalous cases are not completely separated and there is an overlap between them. The log-likelihood values of the AP Overload, Noise and AP Shutdown/Halt scenarios are approximately in a similar range. However, those of the Flash Crowd scenario are slightly lower than the rest and take a widespread range while the values of the AP Shutdown/Halt scenario are condensed in a limited range. 

As a conclusion, all the anomalous cases obtain log-likelihood values less than the normal range and thus it is feasible to distinguish the anomalous time-series from the normal ones. However, due to the overlapping log-likelihood values of the anomalies, it is not that simple to make a distinction between the anomalous scenarios just by inspecting their log-likelihood values. In the next section we consider modeling the anomalous cases independently to facilitate the distinction process. 

\subsubsection{Recognition of anomalous patterns}

\begin{figure}[!t]
\centering
\includegraphics[width=0.49\textwidth]{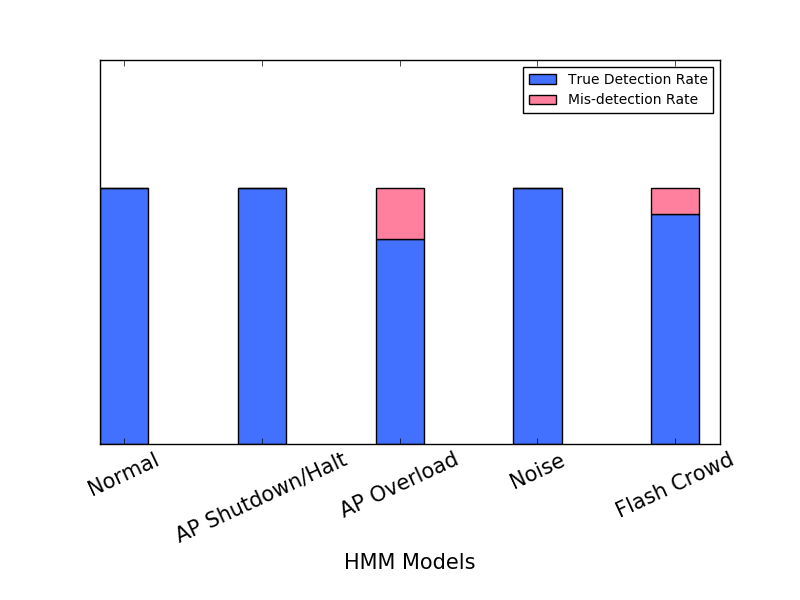}
\caption{Detection results of the observations data by the trained HMM models.}
\label{fig:100}      
\end{figure}

To capture distinctive characteristics of the anomalous scenarios we build separate HMM models for each one and also one model for the normal scenario. Then we compute the probability of each observation sequence getting generated by each of these models. The HMM model that produces the highest log-likelihood value is considered to be the generative model of the given time-series. 

Choosing the best $\lambda$ model among the competing models is termed as \textit{scoring problem} and is an important function of log-likelihood values. At the end of this process we obtain a 2D matrix whose rows and columns consist of HMM models and observation sequences, respectively. 

Figure \ref{fig:100} presents the detection results of the HMM models given the normal and anomalous observation sequences. The x-axis contain the trained HMM models and the bottom part of the bars (in blue) demonstrate the percentage of time-series correctly detected by their corresponding models. The top piece of the bars (in pink) show the mis-detection ratio that occurs in AP Overload and Flash Crowd scenarios. 25\% of AP Overload time-series are detected to be generated by Flash Crowd model. Moreover, 12.5\% of Flash Crowd sequences are detected to be created by AP Overload model and 12.5\% of them by Noise model. Besides these trivial mis-detection errors, the distinction process yields promising results in recognition of different anomalous patterns. 

However, it should be taken into consideration that each anomalous time-series in our work contain a single anomaly, while in reality each time-series can contain no anomaly (in normal cases) or various anomalies (in anomalous cases). A methodology to detect anomalous periods and distinguish between different anomalous patterns in unlabeled data is required to be performed in an unsupervised mode. Here we propose the basic scheme of an algorithm which is based on the general model training in \cite{Ref69}, and is adapted to our specific modeling approach and requisites: 

\begin{enumerate}
  \item A general HMM model is estimated with a large number of training samples (HMM-UBM).
  \item Slice the first test sequence into fixed length segments. The segment(s) with the lowest log-likelihood given the general model in 1 is identified as anomaly.
  \item A new anomalous model is adapted from the general model using the detected anomaly. A normal model is adapted from the general model using the other segments.
  \item Slice the next test sequence into fixed length segments. Estimate the log-likelihood values of all segments given the previous adapted models (normal and anomalous models of step 3). 
  \item Update the adapted models using those segments that achieve closer log-likelihood to each model. Adapt a new anomalous model from the general model using any segment that achieves extremely low log-likelihood given the existing models (a new anomaly that hardly belong to any previous model).
  \item Repeat step 4 and 5 until there is no more test sequences. 
\end{enumerate}

There are a number of parameters in this algorithm that is to be learned and determined, for example the length of the fixed-size segments, and the proper threshold for anomaly detection. However, by the end of this algorithm we expect to have one normal model and several anomalous models each presenting a specific anomalous pattern. Further post-processes are also applicable to merge the very similar models (by measuring models' distance) and yield the most optimized set of final models. More accurate explanation and implementation of this algorithm is out of the scope of the current paper and is left for the future work.

\subsubsection{Detection of anomalous points within a given time-series}

In this approach the anomaly score (log-likelihood) is computed for each data point given the trained HMM model. The unexpected low log-likelihood values show the divergence from the normal model and are typically indicative of anomalies. This method localizes the anomalous points or subsequences more precisely in the test sequence.

To detect the anomalous points in the log-likelihood series automatically, we propose a technique called \textit{threshold detection} to define a boundary where the lower values belong to the anomalous set. 
As many anomaly detection algorithms presume, outliers are the minority group not following the common pattern of the majorities. Accordingly we look for the extreme data points (outliers) with the lowest log-likelihood values. To this end a univariate histogram is constructed and the relative frequency (height of the histogram) is computed. The frequency of samples falling into each bin is used as an estimate of the density. We assume the samples with the highest density (mode) are the normal data points, and accordingly the bins containing the lowest frequencies and farther from the mode are the outliers. As a rule of thumb we mark bins with frequencies lower than a quarter of mode as outliers. Like any other change detection algorithm ours as well produce false positives, however in all the performed experiments of this work the false positive ratio is insignificant. 

We use the same algorithm to detect the outliers or anomalies in RawData and PCA for the purpose of comparison. However, as RawData contains seven features, we conduct the algorithm on each single feature and aggregate the detected points as the final outcome. For example for the likelihood series of $ s_{1}s_{2}...s_{40} $, the algorithm detects $ s_{2} $ and $ s_{4} $ as outlier points for the first feature and $ s_{4} $ and $ s_{15} $ for the third feature and for the rest of the features no anomaly is detected. In this case the final anomalous set contains $ \{s_{2},s_{4},s_{15}\} $. The same method is applied to the PCA components to detect the anomalous points for three principal components.
 
\begin{figure}
\centering
\includegraphics[width=0.48\textwidth]{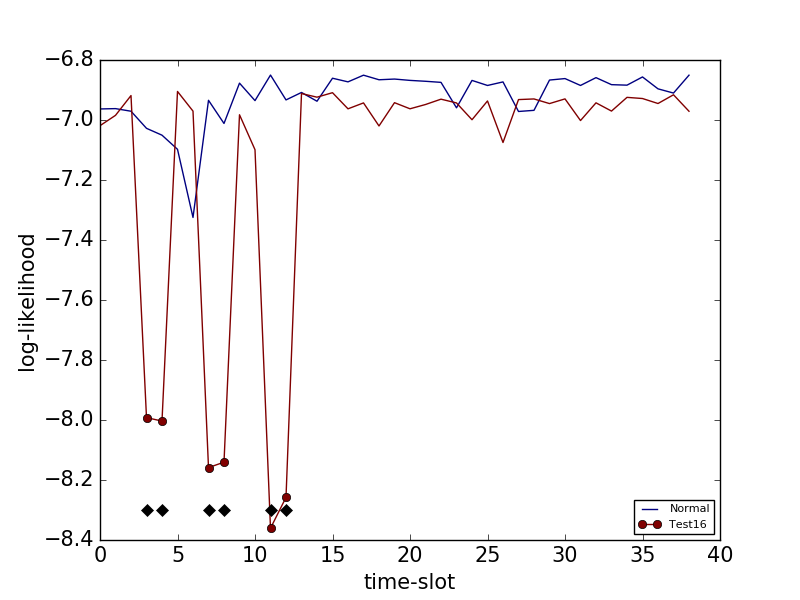}
\caption{Log-likelihood of the normal model together with an example anomaly related to AP Overload experiment.}
\label{fig:1}      
\end{figure}

Figure \ref{fig:1} demonstrates the log-likelihood values of an example anomalous case (AP Overload) generated by simulation. The red points are the anomalies detected by \textit{threshold detection} algorithm and the black diamond points show the real anomalous period. 

We further explore this type of anomaly detection in Section \ref{anomalydetection} and analyze each anomalous case specifically in more detail. 

\section{Experimental Setup}

In order to evaluate the proposed strategy, we perform an extensive set of simulations using OMNeT++ \cite{Ref70} simulator and INET framework \cite{Ref71}. OMNeT++ is a C++-based discrete event simulator (DES) for modeling communication networks, multiprocessors and other distributed or parallel systems. It has a generic architecture and is used in various problem domains including the modeling of wired and wireless communication networks. 

One of the major network simulation model frameworks for OMNeT++ is the INET Framework that provides detailed protocol models for TCP, IPv4, IPv6, Ethernet, Ieee802.11b/g, MPLS, OSPFv4, and several other protocols. We used OMNeT++ along with INET Framework to simulate the IEEE 802.11 WLANg (2.4 GHz band) in infrastructure mode.

In a discrete event simulator, as well as the OMNeT++, events take place at discrete instances in time, and they take zero time to happen. It is assumed that nothing important happens between two consecutive events. Thus the simulation time is relevant to the order of events in the events' queue, and it could take more than the real CPU time or less than it based on the number of nodes, amount of traffic transfered, and other details of the network. 
In our example, with the current number of nodes (5 APs and 30 STAs) and traffic plan, 10 minutes of simulation time takes around 17 minutes of CPU time. Our HMM approach operates on 40 consecutive time-slots of 15s simulation time each.

\subsection{Normal Scenario}
\begin{figure*}[!t]
\subfloat[\label{fig:2-1}]
  {\includegraphics[width=.48\linewidth]{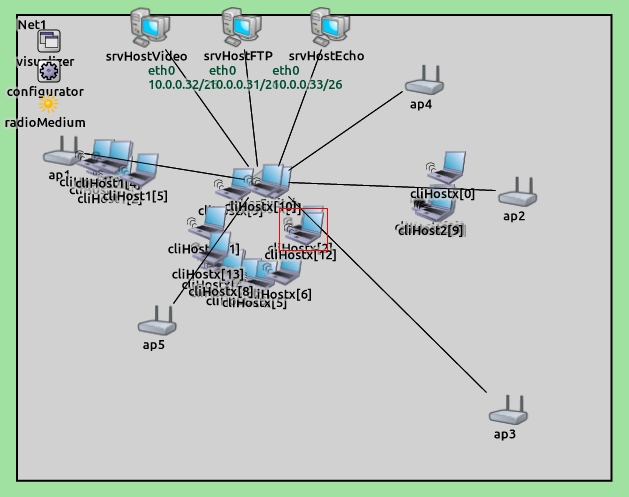}}\hfill
\subfloat[\label{fig:2-2}]
  {\includegraphics[width=.48\linewidth]{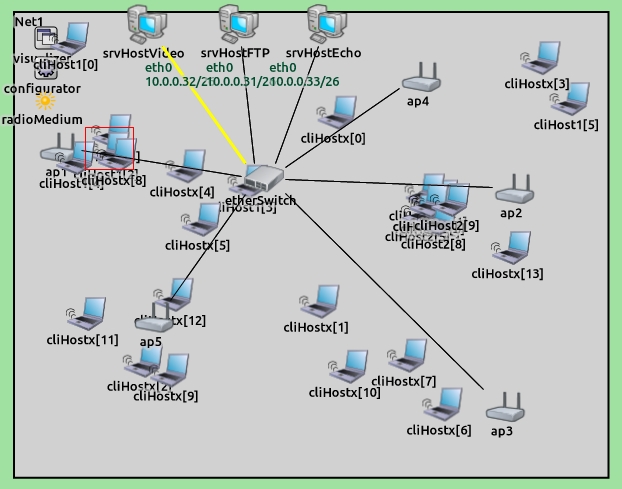}}\hfill
\caption{a) The initial picture of the wireless network simulated in OMNeT++/INET. b) Location of the wireless stations after 30s of simulation.}
\label{fig:2}
\end{figure*}

Figure \ref{fig:2-1} shows the initial picture of a normal scenario, the location of the access points (APs), wireless stations (STAs), and the servers. Figure \ref{fig:2-2} displays the location of the wireless stations after passing 30s (simulation time) from the beginning of the simulation. 

%
%
In the normal scenario, there are 5 APs and 30 STAs. Each STA is initially associated to one of the available APs depending on its location. During the simulation STAs, based on their mobility models, are handed over to other APs when moving around the simulation ground. Furthermore, according to the defined traffic plans in section \ref{traffic}, each node sends and receives packets to the existing servers. 

\subsection{Mobility Models of the Wireless Stations}
The APs are stationary and the wireless nodes follow different mobility patterns. In the current experiment, the mobility models of the nodes are selected in a way to emulate the usage behavior of three typical places in a campus. 
The mobile nodes initially connected to the first AP (AP1) follow the \textit{Linear Mobility} pattern which is configured with speed, angle and acceleration parameters. The mobile nodes move to random destinations with the specified parameters and when they hit a wall they reflect off the wall at the same defined angle. These nodes connect to the other AP (AP2) besides their own AP (AP1), and sometimes lose the connection when they move to blind spots. This pattern is selected to symbolize the nodes with some degree of freedom but within a limited space like administrative offices.

The nodes connected to the second AP (AP2) follow the \textit{Mass Mobility} model, and accordingly move within the room. This pattern of mobility is intended to represent places like classroom or library in which users do not leave the place frequently, but still have some motions in the place. 

The rest of the wireless nodes follow the \textit{Random Waypoint Mobility} and move to a random destination (distributed uniformly over the playground) with a random speed. When the node reaches the target position, it waits for a specified waitTime and selects a new random position afterwards. This type of movement resembles the random mobile users around the wireless ground mostly connected with their mobile devices. 

A summary of wireless nodes' specifications in terms of mobility models is provided in Table \ref{tab:1}.

\subsection{Traffic Generation}
\label{traffic}
As it is shown in Figure \ref{fig:2-1} and \ref{fig:2-2}, there are three main servers wire connected to the Ethernet switch: srvHost\-Video, srvHostFTP, and srvHostEcho. The traffic transfered between wireless stations and the servers (through APs) is considered to be User Datagram Protocol (UDP). The video server (srvHostVideo) sends UDP packets with the message length of $ \mathcal{N} (600B,150B)$ to the clients of AP2, resembling the video downloading by those users. The FTP server (srvHostFTP) is to receive the FTP uploads by the clients of AP1 with message length of $ \mathcal{N} (500B,100B)$. In addition to exclusively downloading or uploading, the other server (srvHostEcho) is in charge of sending and receiving traffics to all the users. This traffic pattern represents the common act of email checking and web browsing by the wireless nodes. The echo packets length are configured to be smaller than the previous ones, $ \mathcal{N} (200B,50B)$, indicating a lighter traffic transmission. In \textit{AP Overload} anomalous scenario one more server is added to take care of heavy channel utilization (srvHostBurst), and more detail about that can be found in section \ref{overload}.

\begin{table}
\centering
\caption{Wireless nodes' specifications in terms of mobility models}
\label{tab:1}      
\begin{tabular}{| p{2.0cm} | p{0.7cm} |>{\raggedright\arraybackslash}p{4.5cm} |}
\hline
Mobility Model & \# Nodes & Mobility Parameters \\
\hline\hline
Linear Mobility & 6 & speed: truncnormal\footnote{Generates random numbers from the truncated normal distribution.} (20mps, 10mps)\newline angle: normal\footnote{Generates random numbers from the normal distribution.} (270deg, 90deg)\newline acceleration: 0 \\ \hline
Mass Mobility & 10 & speed: truncnormal(70mps, 50mps)\newline changeInterval: truncnormal(2ms, 0.5ms)\newline changeAngleBy: normal(90deg, 90deg) \\ \hline
Random Waypoint Mobility & 14 & speed: uniform\footnote{Returns a random variate with uniform distribution in the range [a,b).} (50mps,50mps)\newline waitTime: uniform(3s,8s) \\ \hline
\end{tabular}
\end{table}

\subsection{Path Loss Models}
As the signal propagates through space its power density decreases. Path loss might be due to the combination of many effects, such as free-space loss, refraction, diffraction, reflection, and absorption. 
The path loss model computes the power loss factor based on the traveled distance, the signal frequency and the propagation speed. In our experiments we utilized the following four path loss models to increase the complexity of the simulation and make it more realistic:

\begin{itemize}
\renewcommand{\labelitemi}{$\bullet$}
  \item Free Space Path Loss: is the loss in signal strength resulting from a line-of-sight path through free space, with no obstacles nearby to cause reflection or diffraction.
  \item Log Normal Shadowing: is a stochastic path loss model, where power levels follow a lognormal distribution. It is useful for modeling shadowing caused by objects such as trees.
  \item Rician Fading: is a stochastic path loss model which assumes a dominant line-of-sight signal and multiple reflected signals between the transmitter and the receiver. It is useful for modeling radio propagation in an urban environment.
  \item Rayleigh Fading: is the loss in signal magnitude according to a Rayleigh distribution - the radial component of the sum of two uncorrelated Gaussian random variables. It is useful for modeling the effect of heavily built-up urban environments on radio signals. 
\end{itemize}

\section{Experimental Results and Evaluation}
\label{anomalydetection}
In this section we explore a set of anomalous scenarios and describe different cases of each one. Then we present the HMM and HMM-UBM results in anomaly detection and compare them to baseline approaches (RawData and PCA) for evaluation purposes. 

In terms of HMMs, we consider fully connected models (ergodic), continuous observations with Gaussian distributions, and 3 hidden states. The HMMs with 2 states are too simple to capture the diverse characteristics of the locations (APs), while there is not enough variety in day-long sequential data for 4 or higher number of states. 
Each experiment is repeated at least 20 times with different seeds in order to examine the models on miscellaneous samples providing slightly different data. 80\% of the data sequences is used for training the model and 20\% is kept for testing.

\subsection{AP Shutdown/Halt}

\begin{figure*}[!t]
\subfloat[HMM\label{fig:3-1}]
  {\includegraphics[width=.49\linewidth]{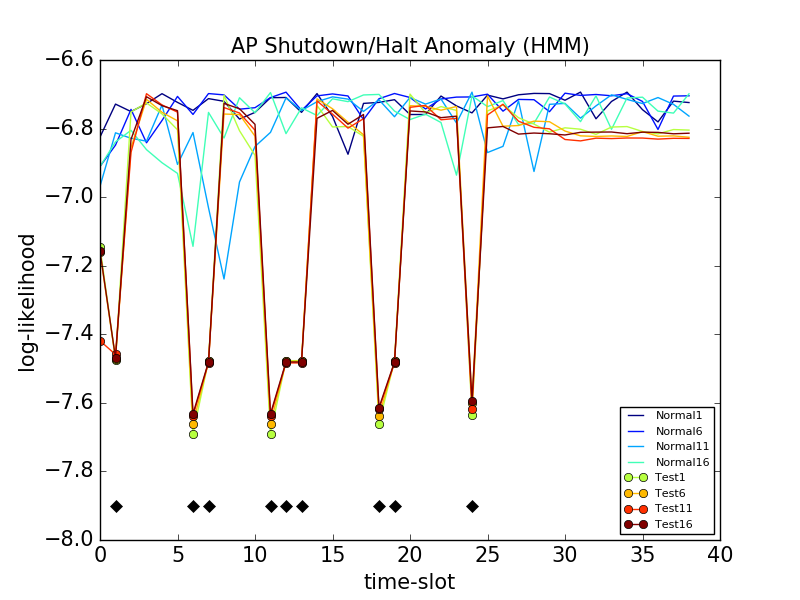}}\hfill
\subfloat[HMM-UBM\label{fig:3-2}]
  {\includegraphics[width=.49\linewidth]{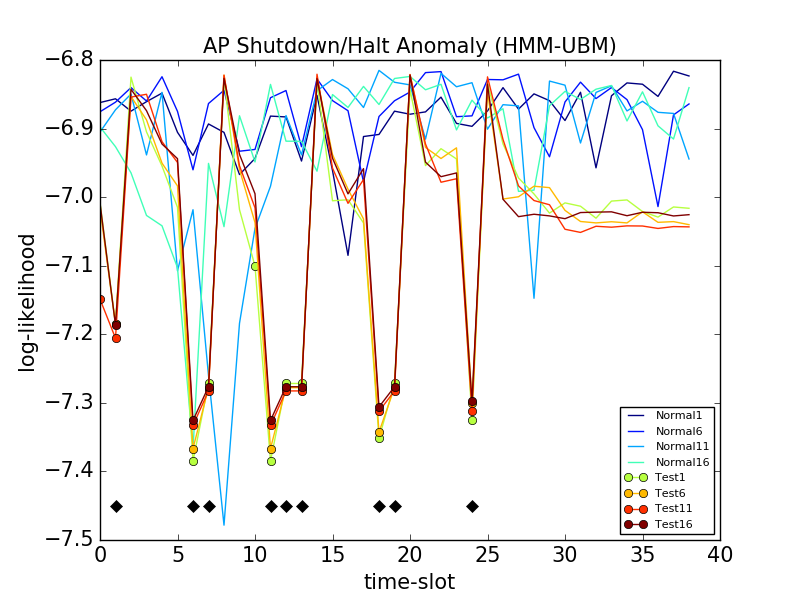}}\hfill
\caption{The log-likelihood series and detected anomalies of AP shutdown/halt scenario in HMM and HMM-UBM models.}
\label{fig:3}
\end{figure*}
\begin{figure}[!t]
\centering
\includegraphics[width=0.49\textwidth]{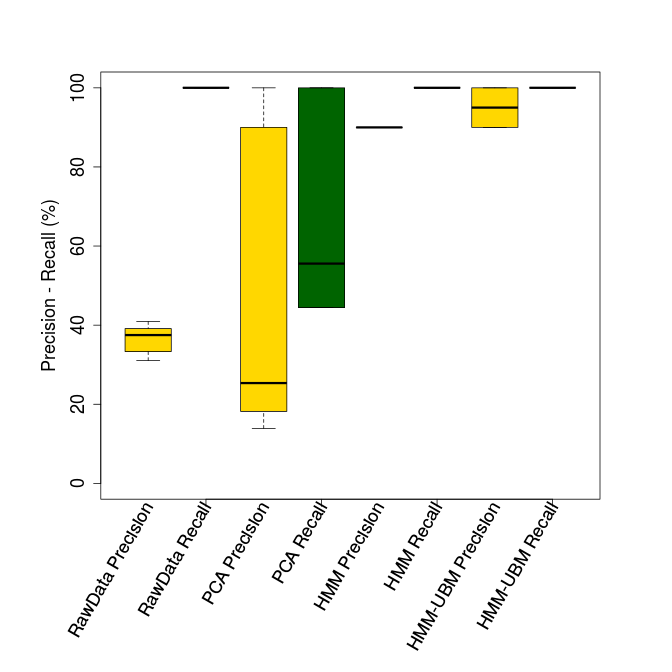}
\caption{Precision and recall boxplot of RawData, PCA, HMM and HMM-UBM belong to AP shutdown/halt scenario.}
\label{fig:4}      
\end{figure}

When there is no session recorded for a given AP in RADIUS accounting table in a period of time, it is likely that the AP has stopped working - possibly due to a technical problem or power failure. In our simulation, we reproduced this anomaly by turning off the AP power deliberately during the \textit{haltperiod} for some \textit{timeslots}. 

Figure \ref{fig:3} demonstrates the HMM likelihood series and the anomalies detected for the test dataset of this scenario. The valley shapes in this image shows the sudden drops of the likelihood values during the anomalous periods, and the marked points are the anomalies detected by the aforementioned \textit{Threshold Detection} algorithm. The black diamonds shows the actual anomalous points generated during the simulation. 

Both HMM and HMM-UBM detect even short shutdown periods that only last for one time-slot. However, Figure \ref{fig:3-1} shows that the HMM model built with only normal data gives a clearer model rather than the HMM-UBM model built with the entire dataset, including the anomalous experiments. Despite this, HMM-UBM obtain great values for precision and recall, and even somewhat higher precision results.

Figure \ref{fig:4} shows the boxplot diagram of the anomaly detection's precision and recall computed for RawData, PCA, HMM, and HMM-UBM models. In these experiments both HMM and HMM-UBM achieve higher precision value and smaller false positive ratio compared to the baseline approaches (RawData and PCA).

Note that this type of anomaly is not very difficult to be detected just by looking at RawData as there is a visible change in dataset features when the power is gone and no session is recorded. That is the reason RawData attains 100\% recall. However, it produces relatively high false positive result that yields low precision.  

\subsection{AP Overload}
\label{overload}

\begin{figure*}[!t]
\subfloat[burstduration $<$ sleepduration\label{fig:5-1}]
  {\includegraphics[width=.33\linewidth]{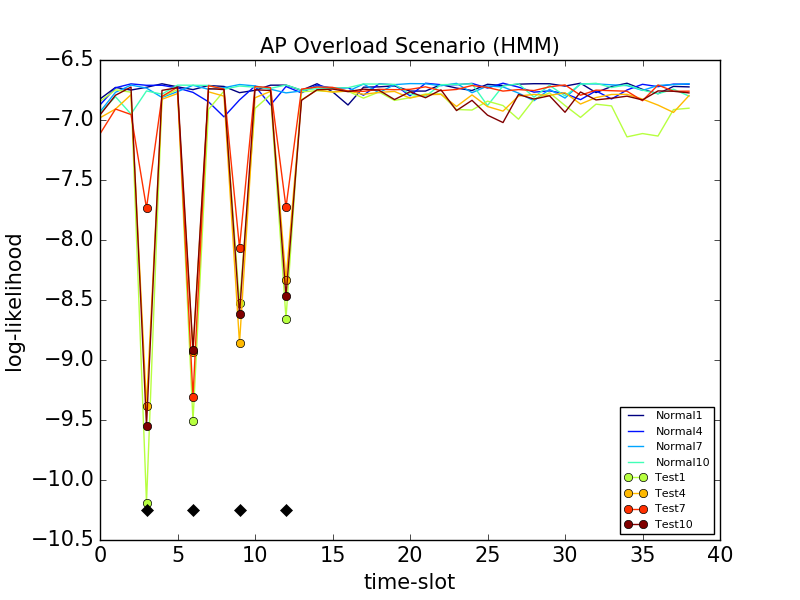}}\hfill
\subfloat[burstduration $=$ sleepduration\label{fig:5-2}]
  {\includegraphics[width=.33\linewidth]{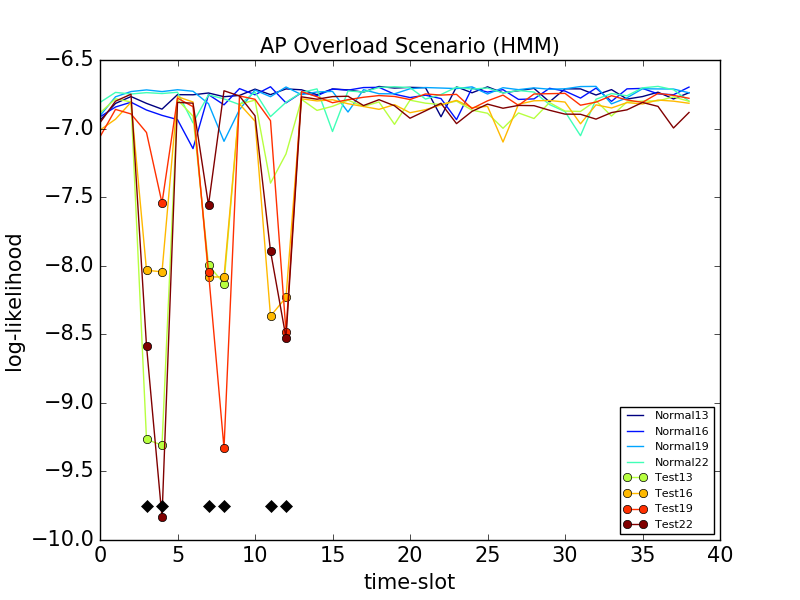}}\hfill
\subfloat[burstduration $>$ sleepduration\label{fig:5-3}]
  {\includegraphics[width=.33\linewidth]{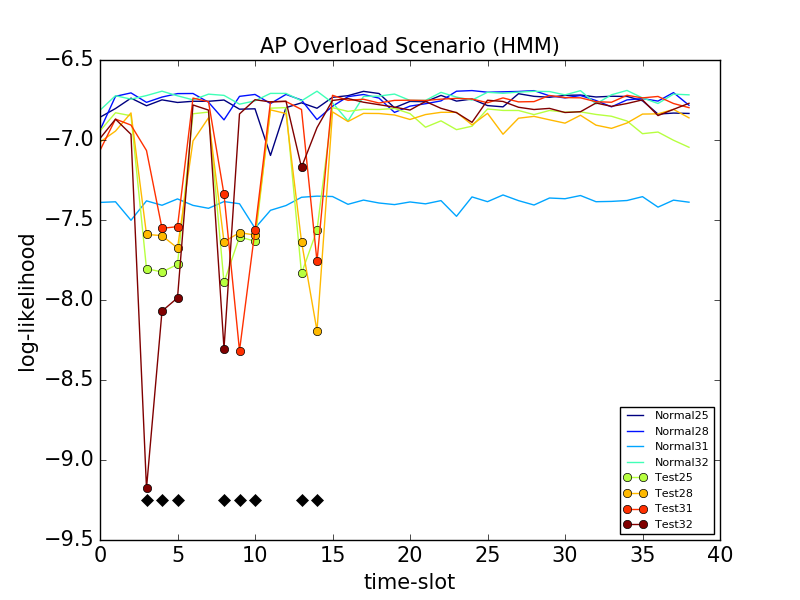}}
\caption{The log-likelihood series and detected anomalies of AP overload scenario (HMM).}
\label{fig:8}
\end{figure*}
\begin{figure*}[!t]
\subfloat[burstduration $<$ sleepduration\label{fig:6-1}]
  {\includegraphics[width=.33\linewidth]{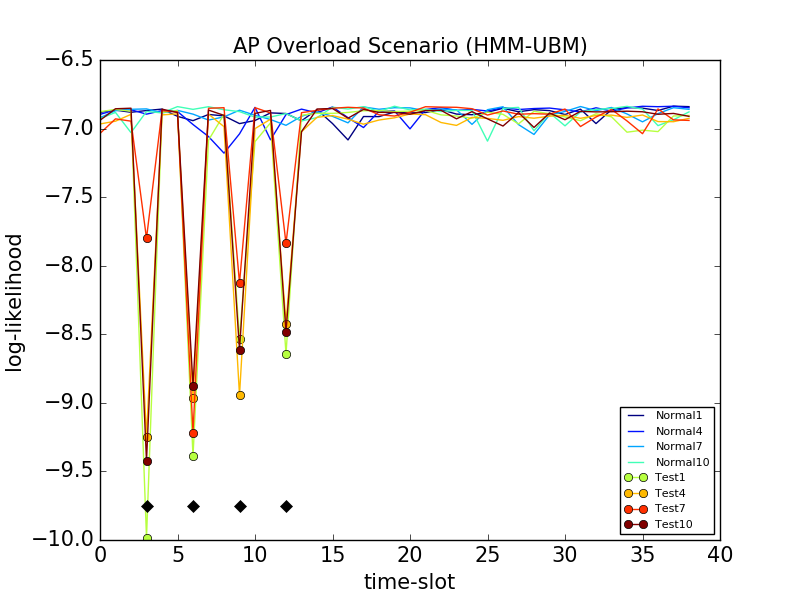}}\hfill
\subfloat[burstduration $=$ sleepduration\label{fig:6-2}]
  {\includegraphics[width=.33\linewidth]{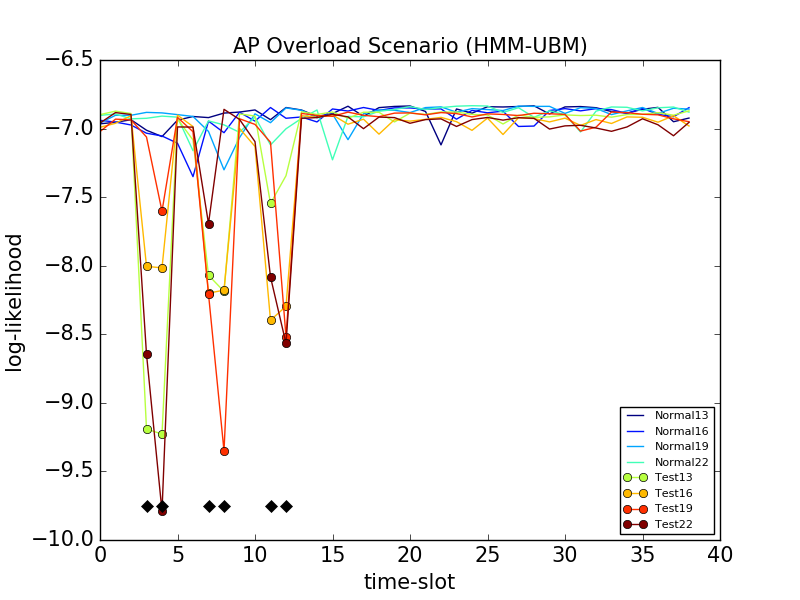}}\hfill
\subfloat[burstduration $>$ sleepduration\label{fig:6-3}]
  {\includegraphics[width=.33\linewidth]{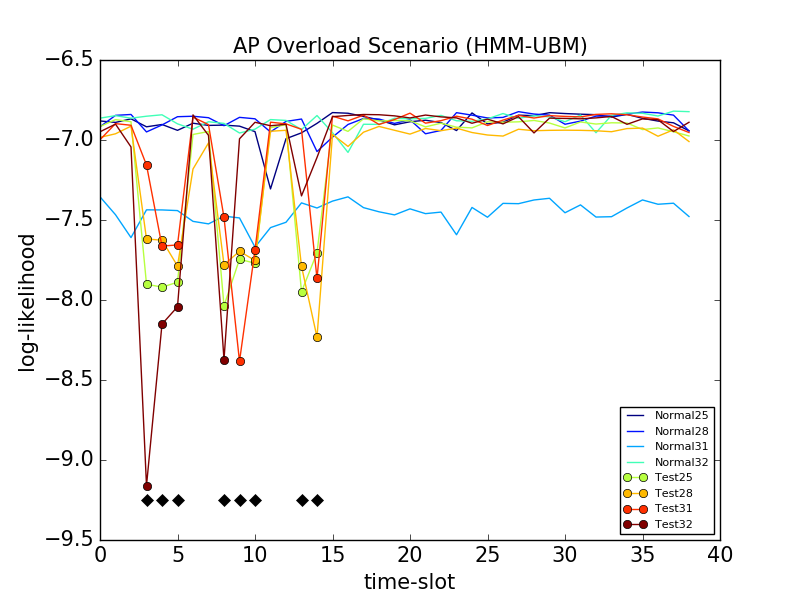}}
\caption{The log-likelihood series and detected anomalies of AP overload scenario (HMM-UBM).}
\label{fig:9}
\end{figure*}
\begin{figure*}[!t]
\centering
\includegraphics[width=0.99\textwidth]{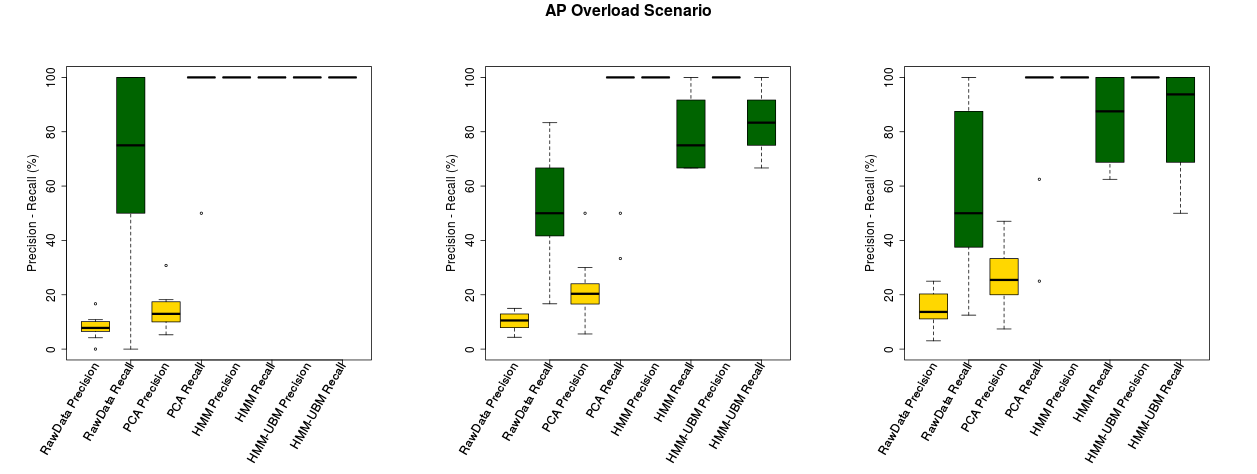}
\caption{Precision and recall boxplot of RawData, PCA and HMM belong to AP overload scenario. Left: burstduration $<$ sleepduration, middle: burstduration $=$ sleepduration, right: burstduration $>$ sleepduration.}
\label{fig:7}      
\end{figure*}

In this anomalous case, the excessive channel utilization occurs that could be the consequence of excessive download or upload by a number of wireless users. In such circumstances, the clients could get disconnected from the current AP frequently even with the presence of high signal strength. In this experiment we simulated AP heavy usage caused by all of the users of the second AP. Burst server (srvHostBurst) sends UDP packets to the given IP addresses in bursts during the \textit{burstduration} which resembles the heavy downloads of the wireless users. In the time of \textit{sleepduration} the burst flow stops and the channel utilization gets back to normal. This experiment contains three different cases as following:

\begin{itemize}
\renewcommand{\labelitemi}{$\bullet$}
    \item burstduration $<$ sleepduration.
    \item burstduration $=$ sleepduration.
    \item burstduration $>$ sleepduration.
\end{itemize}

Figure \ref{fig:8} and \ref{fig:9} display the log-likelihood series of three types of burstduration and sleepduration obtained for AP overload scenario applying HMM and HMM-UBM methodologies, respectively. As it is shown in these figures, during the burst period the log-likelihood value drops drastically and in the sleep period it raises again to the normal level. The longer the burst period the wider is the valley shape in the log-likelihood series, and both HMM and HMM-UBM effectively detect heavy utilization periods in all these cases. 

\begin{figure*}[!t]
\subfloat[-90 dBm\label{fig:14-1}]
  {\includegraphics[width=.33\linewidth]{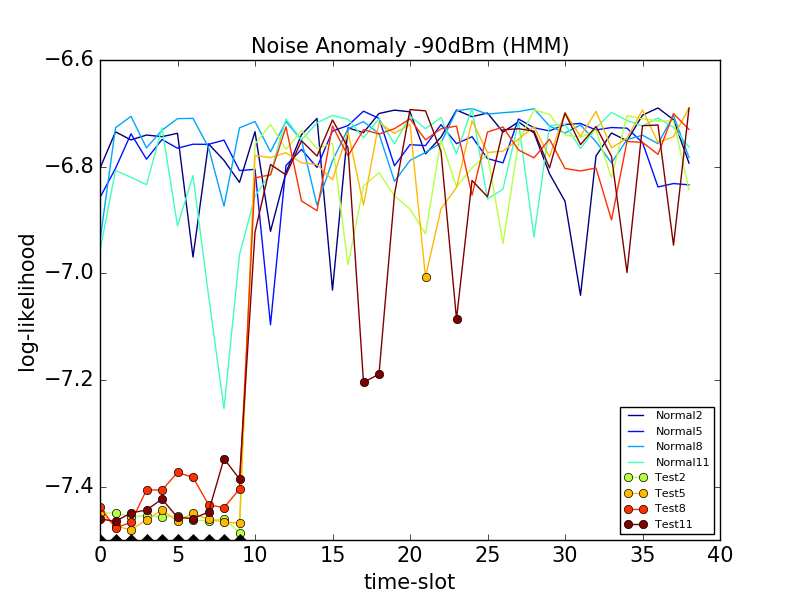}}\hfill
\subfloat[-95 dBm\label{fig:14-2}]
  {\includegraphics[width=.33\linewidth]{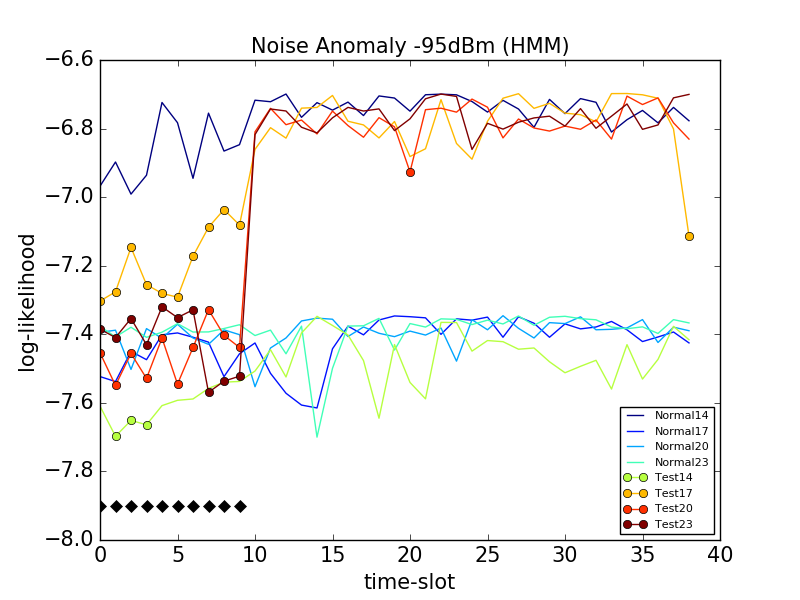}}\hfill
\subfloat[-100 dBm\label{fig:14-3}]
  {\includegraphics[width=.33\linewidth]{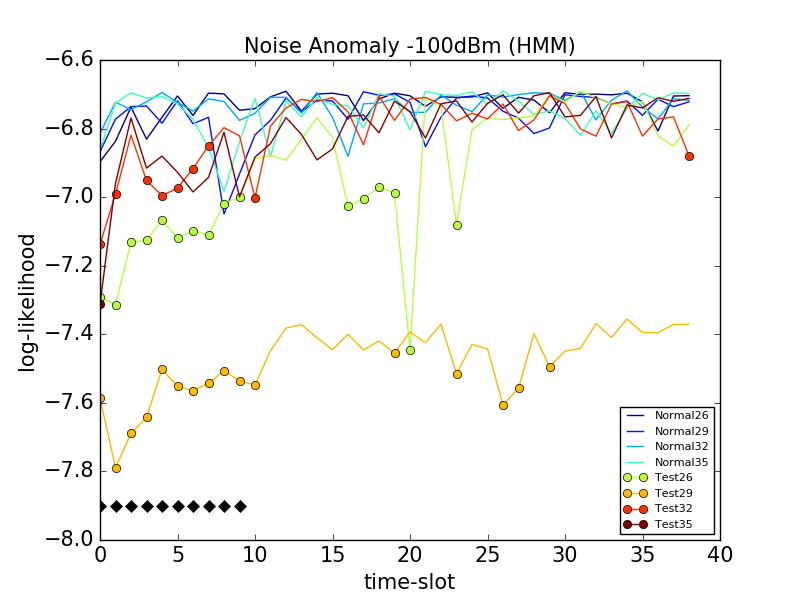}}
\caption{The log-likelihood series and detected anomalies of Noise scenario (HMM).}
\label{fig:14}
\end{figure*}
\begin{figure*}[!t]
\subfloat[-90 dBm\label{fig:15-1}]
  {\includegraphics[width=.33\linewidth]{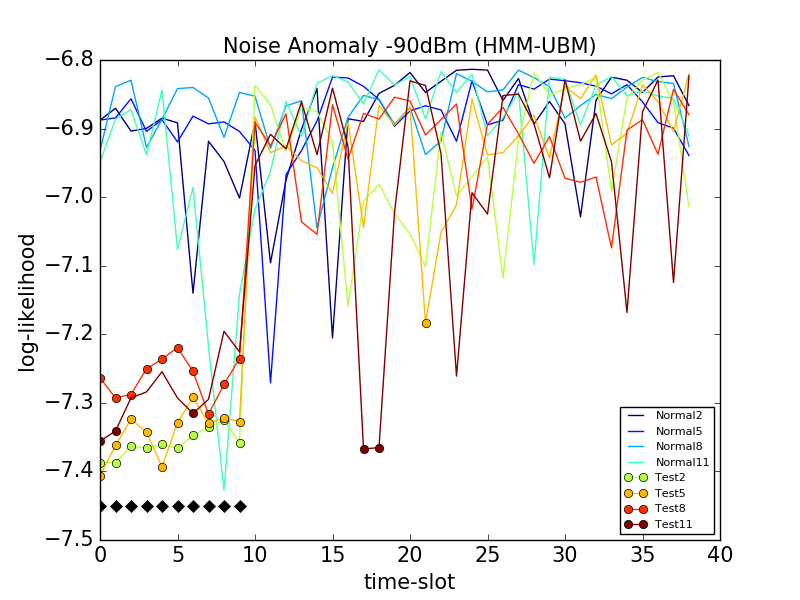}}\hfill
\subfloat[-95 dBm\label{fig:15-2}]
  {\includegraphics[width=.33\linewidth]{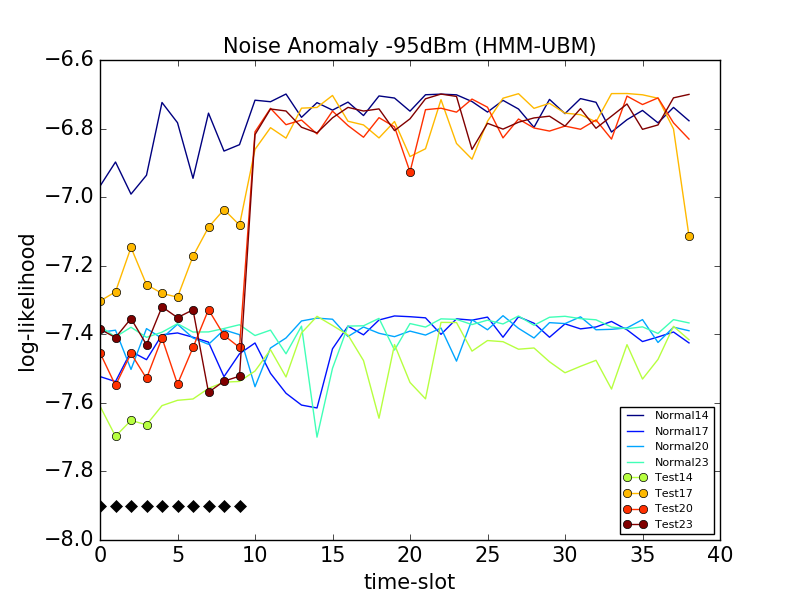}}\hfill
\subfloat[-100 dBm\label{fig:15-3}]
  {\includegraphics[width=.33\linewidth]{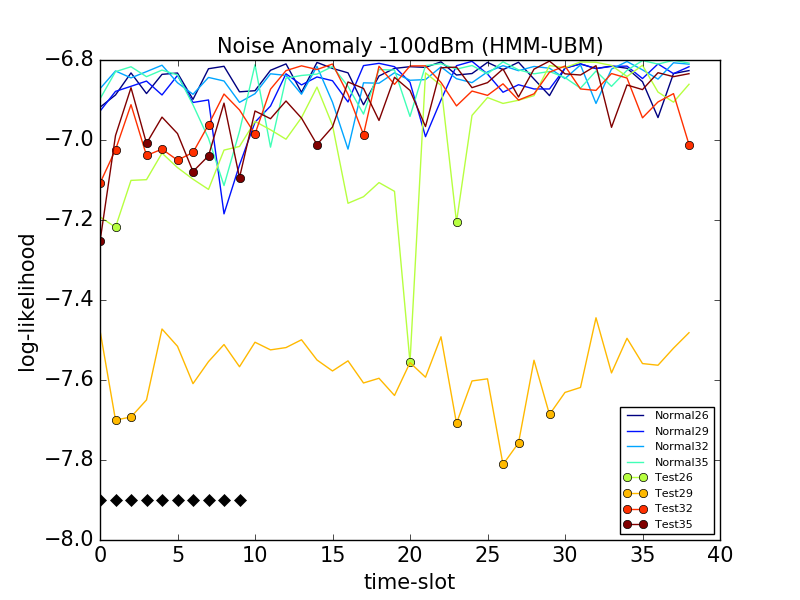}}
\caption{The log-likelihood series and detected anomalies of Noise scenario (HMM-UBM).}
\label{fig:15}
\end{figure*}
\begin{figure*}[!t]
\centering
\includegraphics[width=0.99\textwidth]{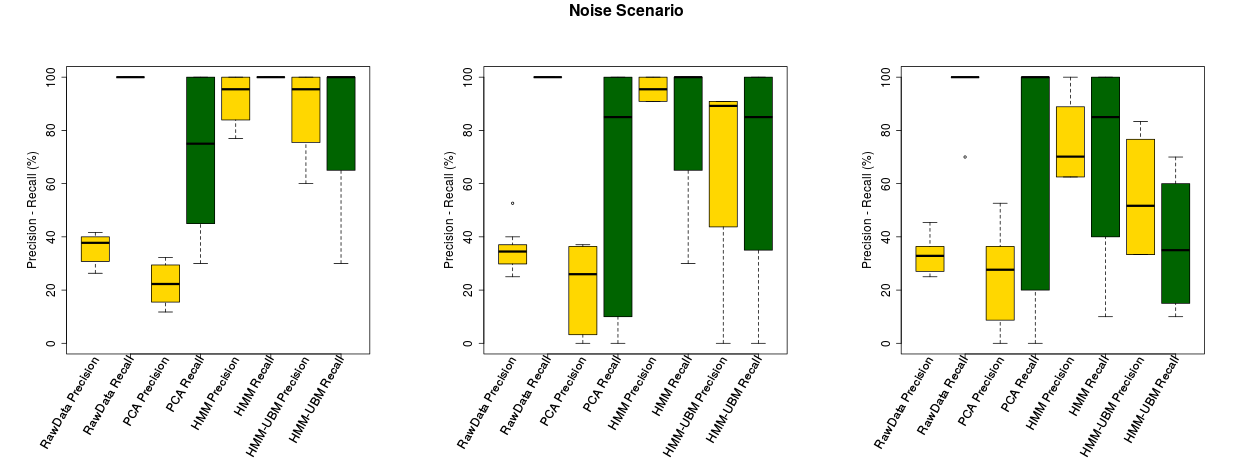}
\caption{Precision and recall boxplot of RawData, PCA, HMM and HMM-UBM belong to noise scenario. Left: -90dBm, middle: -95dBm, right: -100dBm.}
\label{fig:5}      
\end{figure*}

Figure \ref{fig:7} displays the boxplot diagram of the precision and recall results of RawData, PCA, HMM and HMM-UBM models. The low precision ratios of RawData and PCA show that this type of anomaly is not that straightforward to detect directly from the raw data and needs some advanced techniques. The HMM and HMM-UBM results, both in precision and recall, outperform the baseline approaches with a discernible distance. 

\subsection{Noise}

Thermal noise, cosmic background noise, and other random fluctuations of the electromagnetic field affect the quality of the communication channel. This kind of noise doesn't come from a particular source, nor propagate through space. If the noise level is too high, the signal strength will degrade and the performance will decrease.

In the current experiment we change the level of noise power by adjusting the value of \textit{IsotropicBackgroundNoise} parameter in the simulator. The default value of this parameter is set to -110dBm which is the minimum noise level in Wi-Fi networks 802.11 variants. We gradually increase the noise power to -90dBm and record the simulation results repeated 10 times for each experiment. According to the study in \cite{Ref43}, the average noise level in a busy university campus had a stable value at around -94 dBm.

Figure \ref{fig:14} and \ref{fig:15} demonstrate the log-likelihood series of this anomalous scenario, and like previous cases the valley shapes represent the anomalies. The simulated anomalous period is during the first 10 time-slots which is marked with black diamond points. In the first experiment all the anomalous points are detected and the ratio of false positive is quite low. In the next two experiments the detection precision and sensitivity decline. The reason behind this downturn is that as the noise power decreases (higher negative value), it gets more difficult to detect the anomalous periods because the data becomes closer to the normal case (noise power of -110dBm). 

\begin{figure*}[!t]
\subfloat[Flash Crowd Arrival Scenario\label{fig:7-1}]
  {\includegraphics[width=.49\linewidth]{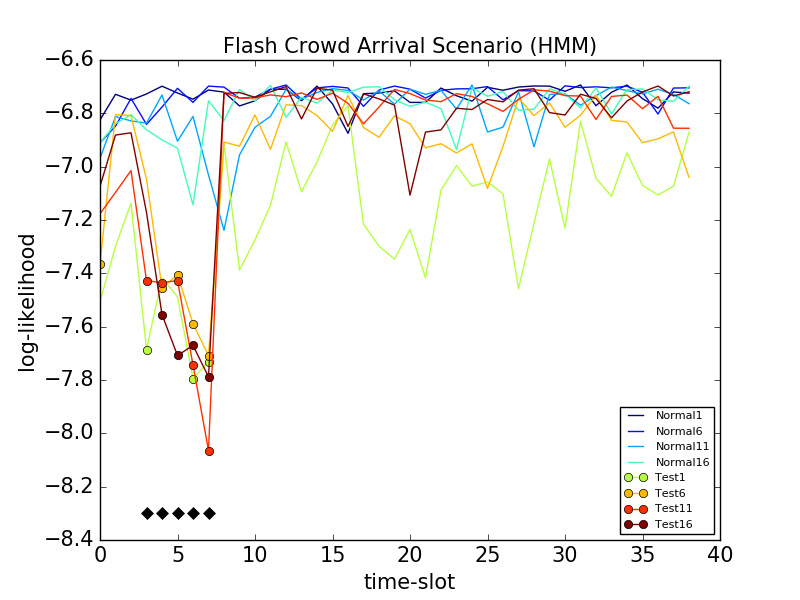}}\hfill
\subfloat[Flash Crowd Departure Scenario\label{fig:7-2}]
  {\includegraphics[width=.49\linewidth]{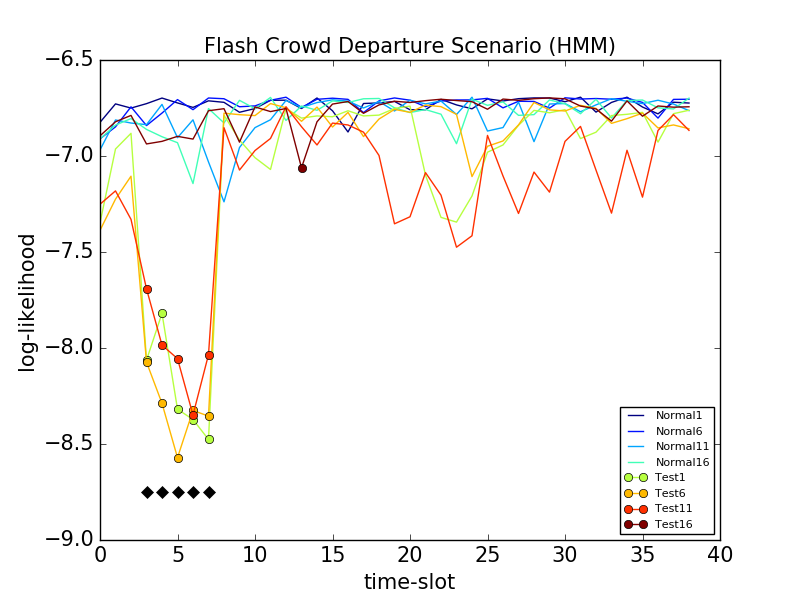}}\hfill
\caption{The log-likelihood series and detected anomalies of Flash Crowd scenario (HMM).}
\label{fig:12}
\end{figure*}
\begin{figure*}[!t]
\subfloat[Flash Crowd Arrival Scenario\label{fig:8-1}]
  {\includegraphics[width=.49\linewidth]{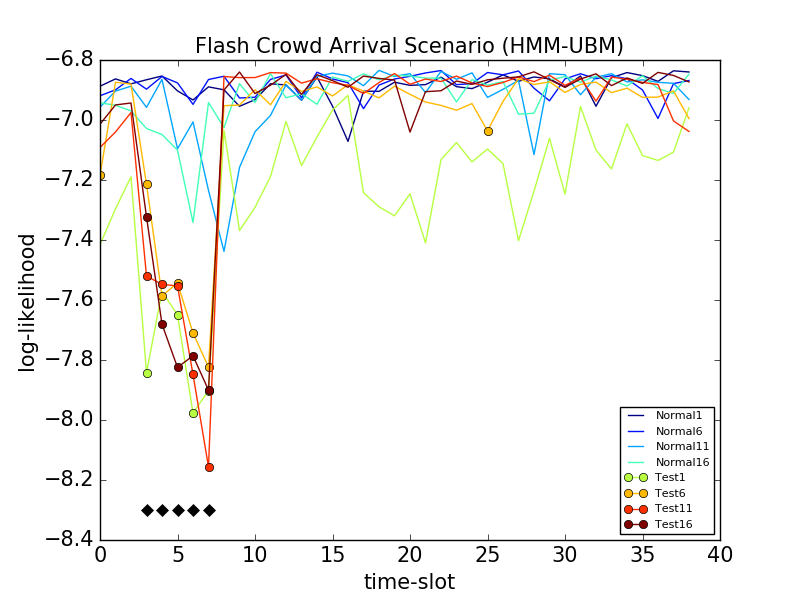}}\hfill
\subfloat[Flash Crowd Departure Scenario\label{fig:8-2}]
  {\includegraphics[width=.49\linewidth]{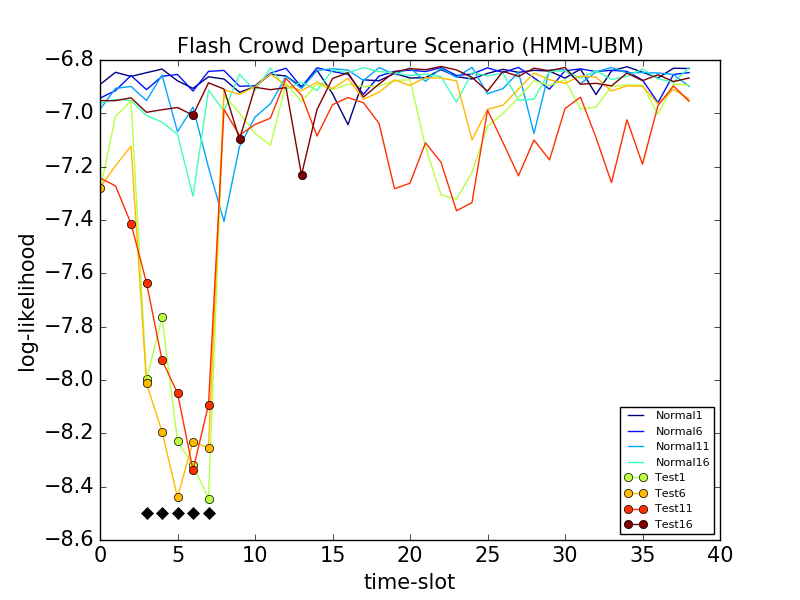}}\hfill
\caption{The log-likelihood series and detected anomalies of Flash Crowd scenario (HMM-UBM).}
\label{fig:13}
\end{figure*}
\begin{figure*}[!t]
\centering
\includegraphics[width=0.95\textwidth]{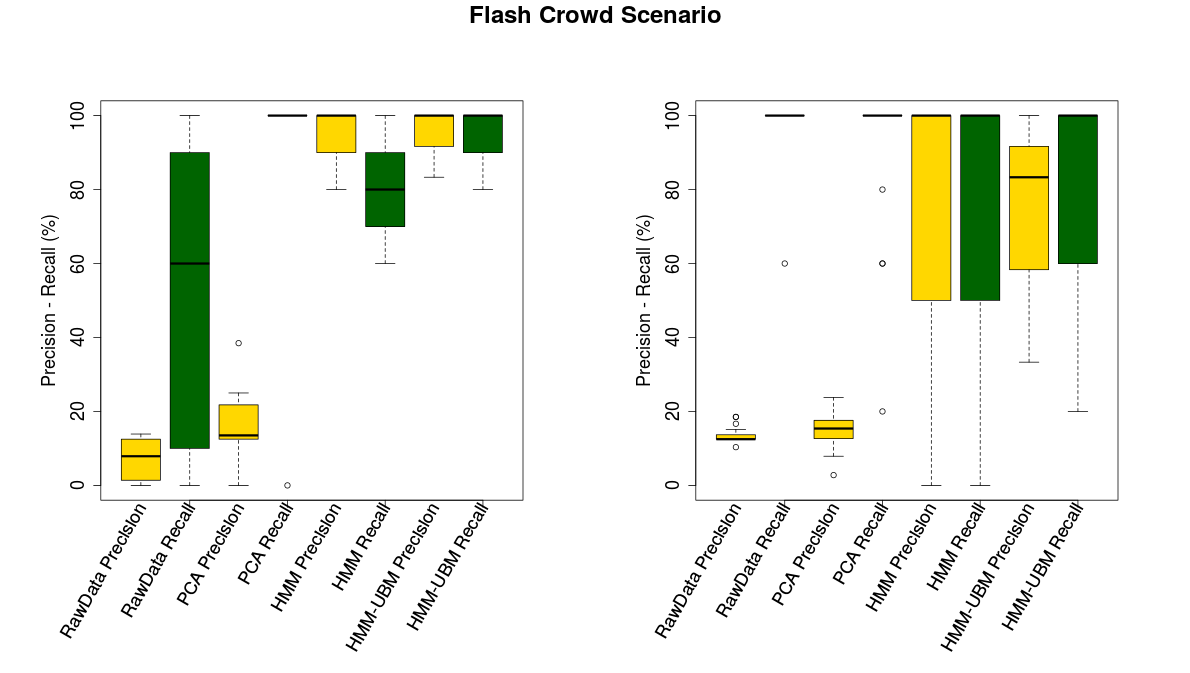}
\caption{Precision and recall boxplot of RawData, PCA and HMM belong to flash crowd scenario. Left: arrival scenario, right: departure scenario.}
\label{fig:11}      
\end{figure*}

As the noise power increases, the packets are less likely to be received at the STAs. Therefore two data features are affected directly by the alteration of noise level: \textit{OutputOctets} and \textit{OutputPackets}. Hence the RawData detector is expected to produce satisfactory detection results. However, as Figure \ref{fig:5} shows, HMM and HMM-UBM models in all the experiments present higher precision values rather than RawData and PCA. 

\subsection{Flash Crowd}

In wireless networks an unexpected surge of traffic occurs mostly due to the beginning or ending of an event when the majority of the wireless users abruptly enter or leave a place and consequently associate to or disassociate from an AP. Such incidents are not necessarily an anomaly in terms of performance or connectivity issues, but could be considered more as a sudden change to a routine network. To see whether the HMM and HMM-UBM model is able to detect such alterations in the normal usage pattern, we simulate this example in two experiments:

\begin{itemize}
\renewcommand{\labelitemi}{$\bullet$}
  \item Arrival: simultaneous association of 7 new nodes to the current AP.
  \item Departure: simultaneous disassociation of 7 existing nodes from the current AP.
\end{itemize}

Figure \ref{fig:13} and \ref{fig:14} present the log-likelihood series of Flash Crowd scenario, and detected anomalous points as colored circles and simulated anomalies as black diamonds. Only in one test case in departure scenario which is related to \textit{Rician Fading} path loss, anomalous period is not detected neither in HMM nor in HMM-UBM. In the rest of the experiments the anomaly detection technique performs accurately both in arrival and departure scenarios. 

As it is illustrated in the boxplot diagram of Figure \ref{fig:11}, HMM and HMM-UBM easily outperform the RawData and PCA results in both \textit{Arrival} and \textit{Departure} scenarios. However, due to the aforementioned exception in the departure scenario, the arrival experiments achieve higher precision and recall.

\section{Conclusions and Future Work}
Intelligent detection of anomalies in 802.11 networks from the analysis of the collected AP usage data is of great significance to network managers. It facilitates their everyday administration workload, and assists them in network maintenance, providing future mitigation plans. 

The key contributions of this work consist of: 1) HMM modeling and threshold detection technique for anomaly detection, 2) proposing HMM-UBM technique for a robust initialization of the hidden states and unsupervised learning, 3) simulation of a small WLAN and a number of anomalous scenarios to evaluate the anomaly detection results.  
 
The precision and recall outcomes of the anomalous cases are computed and compared to the baseline approaches (RawData and PCA). The experimental results show that HMM and HMM-UBM models are both capable of detecting a great portion of anomalies while producing only a trivial false positive ratio. This is promising for in HMM-UBM model all the data, regardless of being normal or containing anomalous events, is utilized to initialize the HMM model. Thus, in unsupervised learning, when the normal data is not known beforehand, HMM-UBM yields a robust model as reliable as HMM for anomaly detection purposes.   
  
In future work we intend to propose a hybrid HMM model that consider the spatial proximity of APs in addition to the temporal relativity of data sequences. Furthermore, we intend to propose an unsupervised learning algorithm for modeling and characterizing various anomaly-related patterns.

\section{Acknowledgement}
This work is financed by the ERDF - European Regional Development Fund through the Operational Programme for Competitiveness and Internationalisation - COMPETE 2020 Programme within project POCI-01-0145-FEDER-006961, and by National Funds through the FCT - Funda\c{c}\~{a}o para a Ci\^{e}ncia e a Tecnologia (Portuguese Foundation for Science and Technology) as part of project UID/EEA/50014/2013. The first author is also sponsored by FCT grant SFRH/BD/99714/2014.

\bibliographystyle{iet}
\bibliography{IETbib}

\newpage

\begin{small}
\noindent {{\bf Anisa Allahdadi} received the B.Sc. in Computer Science from BIHE University (Bah\'{a}'\'{i} Institute for Higher Education), Iran in 2006 and M.Sc in Software Engineering from BIHE University, Iran in 2010. She is currently a researcher in the Center for Telecommunications and Multimedia at INESC TEC and pursuing her Ph.D in the MAP-i Doctoral Programme in the Faculty of Engineering of University of Porto. Her research interests include network management, data mining, probabilistic modeling, and anomaly detection in IEEE 802.11 based wireless networks.}
\end{small}

\vspace{5 mm}
\begin{small}
\noindent {{\bf Ricardo Morla} is an assistant professor of electrical and computer engineering at the University of Porto and principal investigator at INESC TEC, Portugal. His research interests are in the area of automatic system management with an emphasis on probabilistic modeling, prediction, anomaly detection, and root-cause analysis for ICT systems including network infrastructure and smart environments. He holds a Ph.D. in computer science from Lancaster University UK.}
\end{small}

\vspace{5 mm}
\begin{small}
\noindent {{\bf Jaime S. Cardoso} is an Associate Professor with Habilitation at the Faculty of Engineering of the University of Porto (FEUP), where he has been teaching Machine Learning and Computer Vision in Doctoral Programs and multiple courses for the graduate studies. He is a Senior Researcher of the 'Information Processing and Pattern Recognition' Area in the Telecommunications and Multimedia Unit of INESC TEC. He is also Senior Member of IEEE and co-founder of ClusterMedia Labs, an IT company developing automatic solutions for semantic audio-visual analysis. His research can be summed up in three major topics: computer vision, machine learning and decision support systems. He holds a Licenciatura (5-year degree) in Electrical and Computer Engineering in 1999, an MSc in Mathematical Engineering in 2005 and a Ph.D. in Computer Vision in 2006, all from the University of Porto.}
\end{small}

\end{document}